\documentclass[twocolumn,showpacs,preprintnumbers,amsmath,amssymb]{revtex4}

\usepackage{graphicx}
\usepackage{dcolumn}
\usepackage{bm}

\begin{document}


\title{Sex is always well worth its two-fold cost.}
\author{Alexander Feigel\footnote{Electronic address: sasha@soreq.gov.il}, Avraham Englander and Assaf Engel}
\affiliation{%
Soreq NRC,\\
Yavne 81800, Israel}


\date{\today}

\begin{abstract}

Sex is considered as an evolutionary paradox, since its evolutionary advantage does not necessarily overcome the two fold cost of sharing half of one's offspring's genome with another member of the population.  Here we demonstrate that sexual reproduction can be evolutionary stable even when its Darwinian fitness is twice as low when compared to the fitness of asexual mutants. We also show that more than two sexes are always evolutionary unstable. Our approach generalizes the evolutionary game theory to analyze species whose members are able to sense the sexual state of their conspecifics and to switch sexes consequently. The widespread emergence and maintenance of sex follows therefore from its co-evolution with even more widespread environmental sensing abilities.

\end{abstract}

\maketitle

Emergence of sexual reproduction implies universally a two-fold cost to Darwinian fitness\cite{Smith197809,Stearns1988}. A sexual individual shares half of the offspring's genome with another member of the population, while an asexual one generates almost exact replicas of itself. As a consequence, sexual reproduction transfers only half of the amount of the individual's genes to the next generation, when compared to parthenogenesis.

The evolutionary driving forces of sexual reproduction discussed so far, lack either the generality or the strength to overcome the two-fold cost of sex, causing therefore a difficulty to explain evolution and maintenance of sex in some species. Sex increases genetic variability by recombination of the parental chromosomes\cite{Goddard2005,Paland2006}, making a population to be more stable against unpredictable threats, either of internal (deleterious mutations\cite{Kondrashov1988,Barton1998}) or external (parasites\cite{Hamilton1980,Lively1987}, fluctuating environment\cite{Otto1997,Colegrave2002}) origins. It also optimizes the evolutionary search for the optimal gene combinations in a single individual (epistasis\cite{Visser2007}). There are organisms, however, that reproduce sexually, do not have high rate of deleterious mutation, lack parasites and live for a long time in permanent environments.

The paradox is not solved by assuming variability as the most important drive for sexual reproduction. In this case, why are species with more than two sexes not found, and why is asexual reproduction still so widespread\cite{Butlin2002,Hayden2008}? The absence of populations with three or more sexes in nature raises also some doubt concerning genetic variability as the main driving force for evolution of sex. For instance, combination of genetic material from more than two parents may be beneficial under evolutionary conditions that strongly promote a genetically diverse population. There is no known case of more than two sexes, even in case of species that perished due to lack of their ability to resist a significant environmental threat.

The paradox of sex can be solved either by presenting a ubiquitous and strong advantage for sex applicable to all sexual species or by demonstrating how it can evolve despite reduction of Darwinian fitness. In general, the high two fold cost of sex significantly complicates efforts to confirm that sexual reproduction increases Darwinian fitness compared to asexual one. We provide an alternative that successfully solves this paradox, demonstrating that sex can be stable even at twice the cost of parthenogenesis.

To discuss the emergence and maintenance of sex, we assume its onset as evolution of abilities to switch between different sexual states and to sense the sexual distribution of the near environment. These skills remain in modern species: attraction between different sexes is ubiquitous for all animals and some species, even vertebrates, are able to change sex as a function of their environment\cite{Munday2006}. We assume that ultimately the development of the switch/sensor complex led to assortative interactions of complementary sexes, with subsequent degradation of the switching ability.

Consider such organism, able to switch between two sexes $F$ and $M$. Initially, the switch/sensor complex is assumed to choose sex $F$ in any situation, leading to a homogeneous asexual population. A sexual population corresponds to individuals who always choose the sexual state that is opposite to the average sex of the environment: $F$ in a $M$ environment or $M$ in a $F$ environment.

The development of sexual reproduction is governed by its evolutionary payoffs, in our case the growth rates $W_{pq}$ of sex $p$ in environment of sex $q$ $(p,q\in\{F,M\})$. The Darwinian fitness $G$ of an individual can be associated with its mean payoff from all possible situations ($F$ in $F$, $F$ in $M$, $M$ in $F$ and $M$ in $M$). It is determined, therefore, by the individual sensor and switch characteristics, since they affect the probabilities to possess a specific sex under particular circumstances. Evolution proceeds by steps; the existing phenotypes (specific sensor and switch characteristics) in a population are replaced by favorable mutants possessing greater fitness due to their optimal sensor and switch abilities. This process culminates in an evolutionary stable population where no mutants can outperform the host phenotype.

The possibility of existence of a stable, sexually reproducing population with a reduced Darwinian fitness (see Box 1) can be explained briefly as follows. The average fitness in a sexual population $G_{s}$ is given by $(W_{FM}+W_{MF})/2$, since an individual is either $F$ in environment of $M$ or $M$ in environment of $F$. The fitness in an asexual population ($F$ only) is $G_{a}=W_{FF}$. The sexual population is evolutionary stable when the fitness of a parthenogenetic mutant $G_{m}=W_{FM}$ is smaller than $G_{s}$. This occurs when  $W_{MF}>W_{FM}$, that may occur even if $G_{a}>G_{s}$. Rigorous mathematical analysis predicts evolutionary stability of sexual populations even in case of two-fold cost for sex, $G_{a}\approx 2G_{s}$.

In this model a phenotype of an individual $h$ is described by its conditional probabilities $\alpha_{ij}$ to choose sex $i$ in environment of sex $j$. For two sexes $F$ and $M$, it comprises two independent evolvable parameters $(\alpha_{FM},\alpha_{FF})$ corresponding to the conditional probabilities to possess sex $F$ in a $M$ and a $F$ environment respectively. They define the entire set of the conditional probabilities $\alpha_{pq}$: $\alpha_{FM}$, $\alpha_{MM}=1-\alpha_{FM}$, $\alpha_{FF}$ and $\alpha_{MF}=1-\alpha_{FF}$. This formalism follows previous works describing behavior by conditional probabilities\cite{Nowak1990,Skyrms2002,Bergstrom2003,Kussell2005,Camperio2008} and allows to account for any sex determining mechanisms\cite{Doorn2007}.

To determine whether a homogeneous population $(\alpha_{FM},\alpha_{FF})$ is sexual or asexual, as well as to evaluate its Darwinian fitness, one should determine the corresponding unconditional probabilities $\Omega_{pq}$ to possess sex $p$ in environment $q$. According to the definition of $\alpha_{pq}$ as a conditional probability to possess sex $p$ in environment $q$, $\Omega_{pq}$ equals $\alpha_{pq}$ multiplied by the unconditional probability $E_{q}$ to be in environment $q$:
\begin{eqnarray}
\Omega_{MM}&=&(1-\alpha_{FM})E_{M},\nonumber \\
\Omega_{MF}&=&(1-\alpha_{FF})(1-E_{M}),\nonumber \\
\Omega_{FM}&=&\alpha_{FM}E_{M},\nonumber \\
\Omega_{FF}&=&\alpha_{FF}(1-E_{M}).
\label{omegaMF1}%
\end{eqnarray}
Assuming mean-field conditions in a population (properties of the environment are defined by the average value of individual properties), the probability to be in a specific environment matches the individual unconditional probability to possess the corresponding sex:
\begin{eqnarray}
E_{M}=\Omega_{MM}+\Omega_{MF}.
\label{EMmean}%
\end{eqnarray}
Eqs. (\ref{omegaMF1}) and (\ref{EMmean}) result in:
\begin{eqnarray}
E_{M}=\frac{1-\alpha_{FF}}{{1-\alpha_{FF}+\alpha_{FM}}}.
\label{EM1}%
\end{eqnarray}
The statistics $\Omega_{pq}$ can be expressed, therefore, through $(\alpha_{FM},\alpha_{FF})$ only.

A population is sexual when it is composed of individuals with $(\alpha_{FM}=1,\alpha_{FF}=0)$ phenotype. This leads to $(\Omega_{MF}=\Omega_{FM}=1/2,\Omega_{MM}=\Omega_{FF}=0)$ meaning that these individuals switch always to the sex that is complementary to the environment. A population will be asexual when it is composed of individuals with  $(\alpha_{FM}=0,0\leq\alpha_{FF}\leq1)$ or with $(0\leq\alpha_{FM}\leq1,\alpha_{FF}=1)$, leading to  $\Omega_{MF}=\Omega_{FM}=0,\Omega_{MM}=1\;or\;\Omega_{FF}=1\;or\;\Omega_{MM}=\Omega_{FF}=1/2$ respectively. The sexual state of a population will be characterized by $R$, the probability to be in the sex opposite to the environment: $R=\Omega_{FM}+\Omega_{MF}$. A population is sexual when $R=1$, and asexual for $R=0$ (see Fig 1).

The evolutionary stable state of a population is determined by the values of the payoffs $W_{pq}$ to possess sex $p$ in environment of sex $q$ (see Methods):
\begin{equation}
W_{pq}=\begin{array}{c|c|c}
  & F & M \\
\hline
F & 1 & c \\
\hline
M & b & 0  \\
\end{array}. \label{SMWtable4}
\end{equation}
These payoffs also define the cost of sex, corresponding to the ratio of fitness in asexual and sexual population $C_{s}=G_{a}/G_{s}=W_{FF}/((W_{FM}+W_{MF})/2)$:
\begin{equation}
C_{s}=\frac{2}{{b+c}}. \label{Cost0}%
\end{equation}
This expression links evolutionary stability to the cost of sex.

Sexual populations $(\alpha_{FM}=1,\alpha_{FF}=0)$ can be evolutionary stable at a up to two-fold cost for sex $C_{s}=2$ (see Fig. 2). In the case of a greater cost, evolutionary stability exists only for asexual populations. Transition from asexual to sexual population, therefore, may occur with two-fold reduction of individual Darwinian fitness. It should be stressed that (\ref{Cost0}) is valid only for $W_{MM}\ll W_{FF}$ (see SM), which applies probably to all sexual species. Greater values of $W_{MM}$ increase the effective cost of sex and, therefore, require stronger advantages for sexual reproduction.

To compare our results with previous studies, one should analyze the factors contributing to payoffs $b$ and $c$. By taking the individual fitness as the amount of genes in progeny only, one obtains that $b=c=0.5$ with the accompanying two fold cost of sex paradox ($C_{s}=2$, see eq. (\ref{Cost0})). In this case asexual mutants are twice more productive, leading to the seeming evolutionary instability of sexual populations. To solve the paradox, various studies have tried to explain why the fitness of sexual reproduction would be higher than parthenogenesis (reducing $C_{s}$ to $<1$), by considering additional phenomena such as parasites\cite{Hamilton1980,Lively1987}, fluctuating environment\cite{Otto1997,Colegrave2002} and deleterious mutations\cite{Kondrashov1988,Barton1998} that affect payoffs and reduce so the cost of sex. We demonstrate that the sensor/switch abilities make sex evolutionary stable under conditions of $b=1$, $c=0$, keeping $C_{s}=2$ (see Fig. 2). This introduces an asymmetry between female and male payoffs but does not require the reduction of the cost of sex, providing so a general and self-consistent solution for the sex paradox.

Three and more sexes can not be evolutionary stable. To show this, we assume a population to be sexual if its members, when placed in a sexually pure environment, choose with equal probability a sex that is different from the environment (This definition is an extrapolation of the predisposition to be a male in a female environment and vice versa to the case of hypothetical multi-parental reproduction. Other mechanisms that are considered as multi-sexual include explicit requirements for two parents\cite{Hurst1992,Parker2004}). Our approach leads to fractional values of the parameters $\alpha_ {pq}$ for more than two sexes, causing the corresponding populations to be evolutionary unstable (for instance in the case of three sexes: $\alpha_{11}=0,\alpha_{21}=\alpha_{31}=1/2$) (see Methods).

This work presents a rationale for maintenance of sex, providing a universally applicable reason for overcoming the two-fold cost of sex barrier. It explains the widespread phenomenon of sexual reproduction by its link to even more frequently occurring sensing abilities. It allows subsuming existing and future explanations in a framework that decouples the specific mechanisms dealing with emergence and maintenance of sex from the two-fold cost issue. The results are based on a novel approach to incorporate communication in evolutionary game theory, which can be extended to a general analysis of evolution of information exchange and intelligence\cite{Feigel2008}.

\section{Methods}

Consider a population composed of a host $h$, characterized by $(\alpha^{h}_{FM},\alpha^{h}_{FF})$ that is challenged by a mutant $m$, characterized by  $(\alpha^{m}_{FM},\alpha^{m}_{FF})=(\alpha^{h}_{FM}+\Delta\alpha^{m}_{FM},\alpha^{h}_{FF}+\Delta\alpha^{m}_{FF})$.
The evolutionary stability of such a population requires that no mutant is fitter than the host:
\begin{equation}
G(m)-G(h)\leq 0, \label{b1Fm0}%
\end{equation}
where $G(k)$ is the fitness of individual $k$.

The fitness is determined by the individual probabilities $\Omega^{k}_{pq}$ to possess sex $p$ in environment $q$, and by the corresponding payoffs $W_{pq}$:
\begin{equation}
G(k)=\sum_{pq}\Omega^{k}_{pq}W_{pq}. \label{b1GA}%
\end{equation}
Following eqs. (\ref{omegaMF1}) and (\ref{EM1}):
\begin{equation}
G(h)=\sum_{pq}\alpha^{h}_{pq}E_{q}W_{pq}, \label{b1GA1}%
\end{equation}
and
\begin{equation}
G(m)=\sum_{pq}\alpha^{m}_{pq}E_{q}W_{pq}, \label{b1GA2}%
\end{equation}
where $E_{q}$ (\ref{EM1}) is defined solely by the host's values of $(\alpha^{h}_{FM},\alpha^{h}_{FF})$ since we assume that the amount of mutants is small.

Following eqs. (\ref{b1GA1}), (\ref{b1GA2}) and (\ref{SMWtable4}), the condition for evolutionary stability (\ref{b1Fm0}) becomes:
\begin{equation}
\Delta\alpha^{m}_{FM}E_{M}c-\Delta\alpha^{m}_{FF}(1-E_{M})(b-1)\leq 0. \label{b1GA3}%
\end{equation}
This expression divides, for each specific population (a point in phenotype space $(\alpha^{h}_{FM},\alpha^{h}_{FF})$), the phenotype space into two semi-planes corresponding to favorable and non-favorable mutations (see Fig. 3).

According to (\ref{b1GA3}), a sexual population $(\alpha^{h}_{FM}=1,\alpha^{h}_{FF}=0)$ is evolutionary stable for $b>1$ and $c>0$. For other payoff values, a population converges to asexual states with $(\alpha^{h}_{FM}=0)$ or $(\alpha^{h}_{FF}=1)$. As shown in Fig. 3, the linear properties of (\ref{b1GA3}) prevent formation of stable points with fractional values of $\alpha_{pq}$ for sexual populations (unless sex $q$ is not present, $E_{q}=0$ and the corresponding values $\alpha_{pq}$ are irrelevant). This remains valid for more than two sexes (see SM for a rigorous proof).

\appendix

\section{Supplementary material}

\subsection{Derivation of normalization}

Reduction of the payoff table:
\begin{equation}
W_{pq}=\begin{array}{c|c|c}
  & F & M \\
\hline
F & W_{FF} & W_{FM} \\
\hline
M & W_{MF} & W_{MM}  \\
\end{array}, \label{SMWtable5}
\end{equation}
to its two parameters form (\ref{SMWtable4}) requires two transformations:
\begin{equation}
W'_{pq}=W_{pq}-W_{MM}, \label{SMtran1}
\end{equation}
and
\begin{equation}
W''_{pq}=W'_{pq}/(W'_{FF}-W'_{MM}). \label{SMtran2}
\end{equation}
Consequently, the parameters $b$ and $c$ in (\ref{SMWtable4}) are:
\begin{eqnarray}
b&=&\frac{W_{MF}-W_{MM}}{{W_{FF}-W_{MM}}},\nonumber \\
c&=&\frac{W_{FM}-W_{MM}}{{W_{FF}-W_{MM}}}. \label{SMbc}
\end{eqnarray}

The transformations (\ref{SMtran1}) and (\ref{SMtran1}) do not affect the stability condition (\ref{b1Fm0}). Taking into account (\ref{b1GA1}), (\ref{b1Fm0}) becomes $\sum_{pq}\Delta\Omega_{pq}W_{pq}\leq 0$:
\begin{eqnarray}
&&\Delta\Omega_{FF}W_{FF}+\Delta\Omega_{FM}W_{FM}+\ldots\nonumber \\
&+&\Delta\Omega_{MF}W_{MF}+\Delta\Omega_{MM}W_{MM}\leq 0. \label{DO10}%
\end{eqnarray}
where $\Delta\Omega_{pq}=\left (\Omega^{m}_{pq}-\Omega^{h}_{pq}\right )$.

Applying the first transformation (\ref{SMtran1}) to (\ref{DO10}) results in:
\begin{eqnarray}
&&\Delta\Omega_{FF}W'_{FF}+\Delta\Omega_{FM}W'_{FM}+\ldots\nonumber \\
&+&\Delta\Omega_{MF}W'_{MF}+\Delta\Omega_{MM}W'_{MM}+\ldots\nonumber \\
&+&W_{MM}\sum_{pq}\Delta\Omega_{pq}\leq 0. \label{b1GB}%
\end{eqnarray}
The last term in the left part vanishes $(\sum_{pq}\Delta\Omega_{pq}=0)$ preserving the form of condition (\ref{b1GA}). The second transformation (\ref{SMtran2}) converts (\ref{b1GB}) into $\sum_{pq}\Delta\Omega_{pq}W''_{pq}\leq 0$.

Expressions for cost of sex (\ref{Cost0}) and the ratio of fitness in asexual and sexual population $C_{s}=W_{FF}/((W_{FM}+W_{MF})/2)$ are identical only in case $W_{MM}=0$. Otherwise, using (\ref{SMbc}):
\begin{eqnarray}
\frac{2}{{b+c}}=\frac{2W_{FF}-2W_{MM}}{{W_{MF}+W_{FM}-2W_{MM}}}. \label{Costcom}%
\end{eqnarray}
Finite positive values of $W_{MM}$, therefore, increase effective cost of sex:
\begin{eqnarray}
\frac{2}{{b+c}}>\frac{2W_{FF}}{{W_{MF}+W_{FM}}}. \label{Costcom1}%
\end{eqnarray}

\subsection{Evolutionary stability for more than two sexes}

In case of more than two sexes, fractional values of $\alpha_{pq}$ can not be evolutionary stable, since the only possibility to confine a point with a line on the edge of a square is when the line is parallel to the edge (see Fig 3, point {\sc\romannumeral 4}). This occurs only if the corresponding sex $q$ is not present in the population ($E_{q}=0$). The same applies to multiple sexes, with $K$ dimensional cube and $K-1$ dimensional constraint for evolutionary stability.

A rigorous proof is as follows. The evolutionary stability of specific value of $\alpha_{ij}=\alpha_{st}$ requires $\partial G/\partial \alpha_{ij}=0$ at $\alpha_{st}$ in case $\alpha_{st}\neq 0$ and $\alpha_{st}\neq 1$. In case of multiple sexes, the expression for fitness remains identical to (\ref{b1GA1}):
\begin{equation}
G=\sum_{pq}\alpha_{pq}E_{q}W_{pq}. \label{SMep2}
\end{equation}
Consequently:
\begin{equation}
\frac{\partial G}{{\partial \alpha_{ij}}}=E_{j}\sum_{i}W_{ij}, \label{SMep3}
\end{equation}
This expression vanishes in case $E_{j}=0$, meaning that sex $j$ is not present in the population.

\subsection{Comparison with model games}

In evolutionary game theory, the payoffs (\ref{SMWtable4}) are separated into standard models games. According to this work, the games of Leader $(b>c,c>1)$, Battle of the Sexes $(b<c,b>1)$ and Chicken $(b>1,0<c<1)$ lead to development of sexual reproduction, while Prisoner's dilemma $(b>1,c<0)$ promotes asexual populations.

\begin{widetext}
\clearpage
\begin{figure}
  \begin{center}
      \resizebox{0.5\textwidth}{!}{\includegraphics{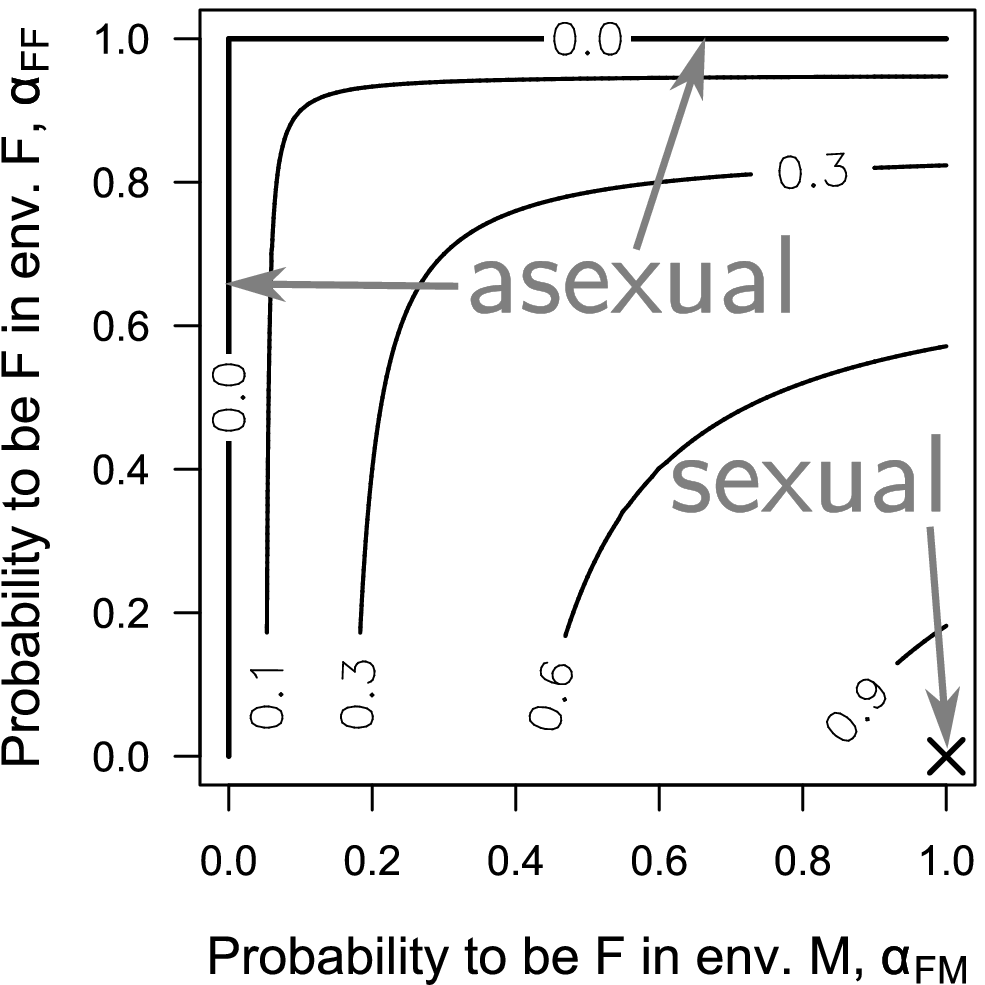}}
    \caption{{\bf Evolution of sexual reproduction as development of sex switching and sensing abilities.} Contour plot of $R$ (probability to possess sex complementary to the environment) in populations with varying individual switch and sensor characteristics. The switch/sensor complex is described by $\alpha_{FM}$ and $\alpha_{FF}$, representing conditional probabilities to possess sex $F$ in $M$ and $F$ environments respectively. The single point $\times$ denotes a fully developed sexual population ($R = 1$) while there exist multiple possibilities for asexual populations ($R = 0$). Evolution is equivalent to the motion of a point, denoting a population, from an asexual state to the sexual endpoint. Specific evolutionary mechanisms correspond to different evolutionary pathways.}
    \label{fig1}
  \end{center}
\end{figure}
\clearpage
\begin{figure}
  \begin{center}
      \resizebox{0.5\textwidth}{!}{\includegraphics{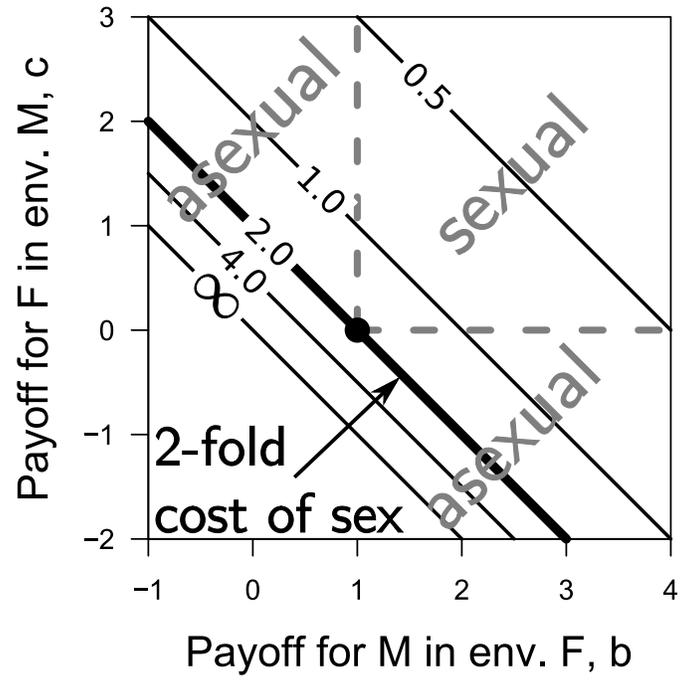}}
    \caption{{\bf Evolutionary stability of sexual population for different values of cost of sex.} The space of evolutionary payoffs, separated into sexual and asexual evolutionary stable regions, superimposed on a contour plot of the cost of sex ($C_{s}$). Many previous works suggested that sex is stable only if its fitness is higher than the parthenogenetic one ($C_{s}<1$ region). We demonstrate that the sensor/switch abilities make sex evolutionary stable for up to two-fold cost ($C_{s}=2$). No stability is possible for $C_{s}>2$.}
    \label{fig2}
  \end{center}
\end{figure}
\clearpage
\begin{figure}
  \begin{center}
      \resizebox{0.5\textwidth}{!}{\includegraphics{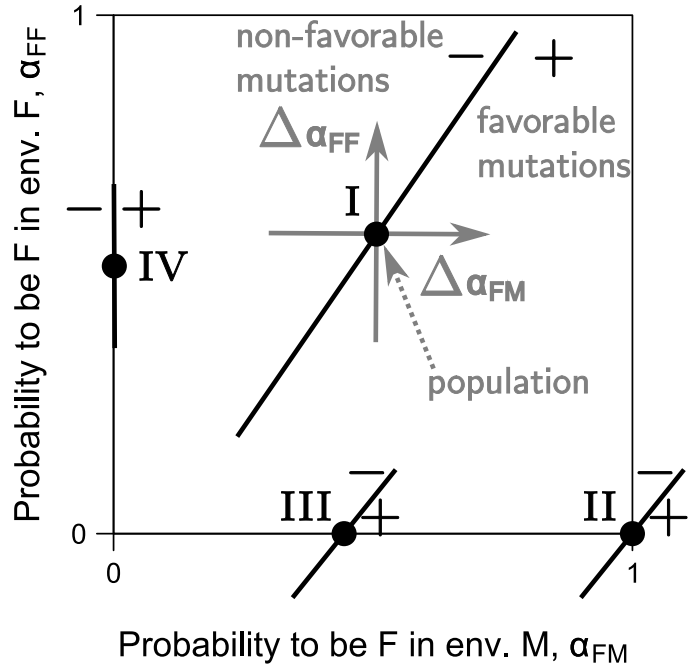}}
    \caption{{\bf Evolutionary stability of sexual reproduction.} The equation for evolutionary stability (\ref{b1GA3}) defines positive and negative semi-planes, corresponding to favorable and non-favorable mutations, dependent on payoffs $b$ and $c$. A population with arbitrary $\alpha_{FM}$ and $\alpha_{FF}$ ({\sc\romannumeral 1}) will always dispose of a positive region, precluding so an evolutionary stable solution. A sexual population ({\sc\romannumeral 2}) is evolutionary stable since no positive direction is available. The population on the $\alpha_{FM}$ ({\sc\romannumeral 3}) and $\alpha_{FF}$ ({\sc\romannumeral 4}) edges are also unstable. If we assume opposite signs for the semi-planes as consequence of different payoff values, the asexual populations (positioned on the edges $\alpha_{FM} = 0 $ or $\alpha_{FF} =1$) become stable, while the sexual population becomes unstable.}
    \label{fig3}
  \end{center}
\end{figure}
\end{widetext}
\end{document}